\documentclass[journal]{IEEEtran}
\usepackage{graphicx} 
\usepackage{cite}
\usepackage{url}
\usepackage{hyperref}
\usepackage{amsmath,amssymb,amsfonts}
\usepackage{algorithmic}
\usepackage{textcomp}
\usepackage{authblk}

\begin{document}
\title{DAWN: Designing Distributed Agents in a Worldwide Network}
\author{Zahra Aminiranjbar$^{*}$, Jianan Tang$^{*}$, Qiudan Wang, Shubha Pant, Mahesh Viswanathan$^{*\dagger}$,~\IEEEmembership{IEEE Fellow} \thanks{$^{*}$ Equal contributions} \thanks{$^{\dagger}$ Corresponding Author: mahviswa@cisco.com}}%
\affil[]{Cisco Systems, Inc.}
\affil[]{\texttt{\{zaminira, jianatan, cathewan, shubpant, mahviswa\}@cisco.com}}
\date{May 2025}
\maketitle

\begin{abstract}

The rapid evolution of Large Language Models (LLMs) has transformed them from basic conversational tools into sophisticated entities capable of complex reasoning and decision-making. These advancements have led to the development of specialized LLM-based agents designed for diverse tasks such as coding and web browsing. As these agents become more capable, the need for a robust framework that facilitates global agentic communication and collaboration for building sophisticated software solutions has become increasingly important. Distributed Agents in a Worldwide Network (DAWN) addresses this need by providing an architectural framework that allows globally distributed agents of any provenance to be registered, discovered, and organized for building AI-based applications and solutions. In DAWN, a Principal Agent Service composes and oversees the execution of agentic applications. It delegates tasks to one or more Gateway Agent Services that provide for the discovery, registration, and connection of the most suitable agents to fit each application's need. DAWN offers three operational modes: No-LLM mode for deterministic and classical software development, Copilot for decision-making augmented using AI, and LLM Agent for autonomous operations. Last but not the least, DAWN ensures the safety and security of agent collaborations globally through a dedicated safety, security, and compliance layer, protecting the network against attackers and adhering to stringent security and compliance standards. These features make DAWN a robust framework for designing, developing and deploying agent-based applications across business and consumer applications.
\end{abstract}

\section{Introduction}
Rapid advancements in large language models (LLMs)\cite{openai2024chatgpt, touvron2024llama3, claude3} have have sparked significant interest in LLM-based AI agents\cite{guan2024intelligent} which have the potential to free humans from repetitive tasks and significantly enhance productivity. New (AI) agents are being developed on a regular basis and their performance is correspondingly improving. Agents exist for tasks such as coding\cite{openinterpreter2024,nijkamp2023codegen}, web browsing\cite{zhou2024webarena}, gaming, autonomous assistants, computer use, and others \cite{wang2024voyager, zhang2024proagent}. Furthermore, multi-agent systems\cite{wuautogen,hong2024metagpt, li2023camel,chen2024agentverse, langgraph2023,openai_swarm,magenticone2024} have been proposed to leverage the collective intelligence and specialized profiles and skills of multiple agents. And agents are gradually finding their way into business processes and software development. However, three major challenges arise that hinder their potential for broader and more effective application.

\begin{figure*}
    \centering
    \includegraphics[width=\textwidth]{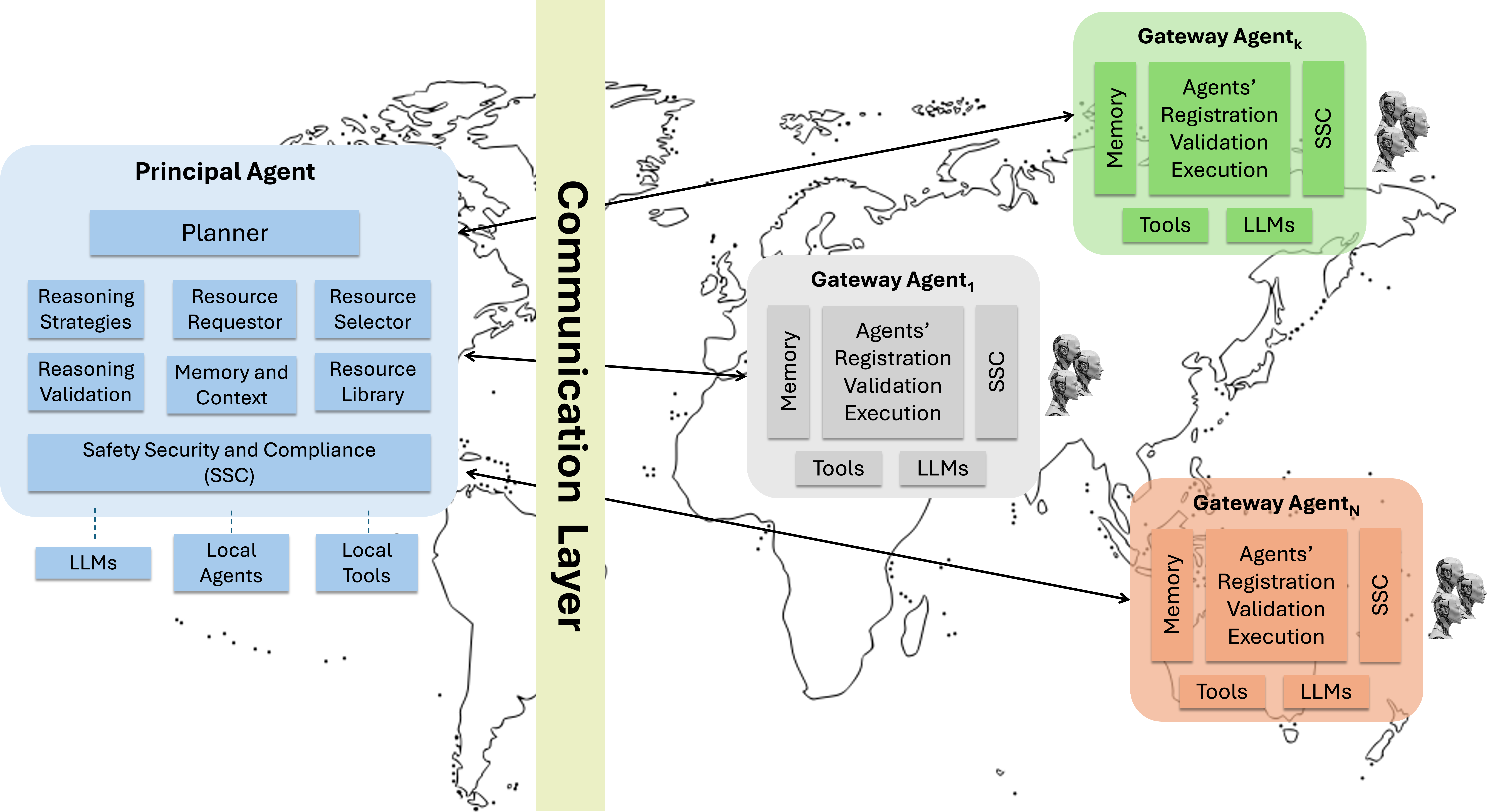}
    \caption{DAWN architecture showing its main components and capabilities.}
    \label{fig:1}
\end{figure*}

The first challenge is how to discover and connect to agents to accomplish tasks when they are distributed globally. Existing frameworks, such as LangGraph\cite{langgraph2023} and AutoGen\cite{wuautogen} have bounded scope as they are typically designed to manage a limited number of agents that work well when created within their frameworks.  This constraint arises from the need to manually hard-code the names, URLs, and description of other agents into the prompts, which becomes increasingly difficult as the network of agents grows. More agents created by different entities allow for a broader selection covering functional and non-functional capabilities. Additionally, integrating heterogeneous agents from different ecosystems requires developers to invest effort in developing wrappers and packages specific to each new framework as they evolve. New frameworks such as Agntcy\cite{cisco_agntcy_2025}, A2A\cite{google_a2a_2025}, and ACP\cite{agentcommunicationprotocol_welcome_2025} mitigate some of the issues relating to finding and connecting agents.

The second challenge arises because many business use cases demand determinism, consistency, and reliability, whereas LLM-based agents are inherently stochastic and excel in more flexible environments. They struggle with tasks requiring strict adherence to predefined rules and execution graphs. Empirical evidence shows that manually designing workflows, rather than granting full autonomy, leads to more robust performance and lower computational costs \cite{xia2024agentlessdemystifyingllmbasedsoftware}. Balancing these often conflicting needs for autonomy and determinism is therefore essential to the broader adoption of agent-based systems. Moreover, agentic software development is expected to be a combination of AI and non-AI tools working together complicating the needs of business processes predictability while exploiting the power of AI.

The third challenge is ensuring the safety and security of agentic applications. LLM-based Agents are vulnerable to specific attacks such as prompt injection \cite{greshake2023youvesignedforcompromising} and jailbreaking \cite{yu2024gptfuzzerredteaminglarge,owasp2025llm}. In a global agentic framework, the use of dynamic capabilities can increase their exposure to widespread risks. For example, when LLM agents use third party APIs\cite{openai2024}, malicious APIs can plan attacks and insert, substitute, or delete key information. Depending on the agent's application domain, such attacks could threaten physical security, financial security, or overall system integrity. This underscores the critical importance of applying robust safety and security measures in such frameworks. Add to this the fact that generalized LLMs are invoked via provider APIs external to most business and consumer networks, and data is pushed to them for analysis. This increases the risk of data exfiltration and concomitant privacy risks. Safety, security, and privacy of business and consumer data needs to be handled with great care. This requires the addition of control points during the design, build, and deploy phases of agentic software development.

The proposed Distributed Agents in a Worldwide Network (DAWN) framework addresses these three issues. DAWN's architecture includes key components such as the Principal Services Agent and Gateway Services Agent - Principal Agent and Gateway Agent, for short. These ensemble services work together to enable globally distributed agents to plan, compose, discover, connect, orchestrate, observe, safeguard, and communicate. Principal Agents are connected with a fixed set of Gateway Agents in a fashion similar to sub-contractors associated with a principal contractor on a manufacturing contract. Given a user request, the Principal Agent autonomously plans and then produces a list of tasks and agentic workflow describing the derived capabilities to subsidiary Gateway Agents. Gateway Agents, which may be local or distributed globally, have the capability to register public and proprietary resources (tools, agents, and agentic applications). Upon request, the Gateway Agents search their registries for the most suitable resources to address the proffered task list. These selected resources are wrapped into RESTful endpoints and returned to the Principal Agent. The Principal Agent then orchestrates these resources across Gateway Agents to execute the plan. Fig. \ref{fig:1} illustrates the globally distributed nature of this collaboration of an Principal Agent and Gateway Agents. As evident from the figure, the Principal Agent and attached Gateway Agents are more than mere single-function AI agents or applications. They host multiple functionalities and capabilities to support the creation, development, deployment, and operation of many applications that may be devised statically (predesigned and predeployed compute resources to address well-known workplace and consumer problems) or dynamically (on-the-fly accumulation and threading of AI agents and related resources along with the requisite compute infrastructure). 

\begin{figure}
    \centering
    \includegraphics[width=1\linewidth]{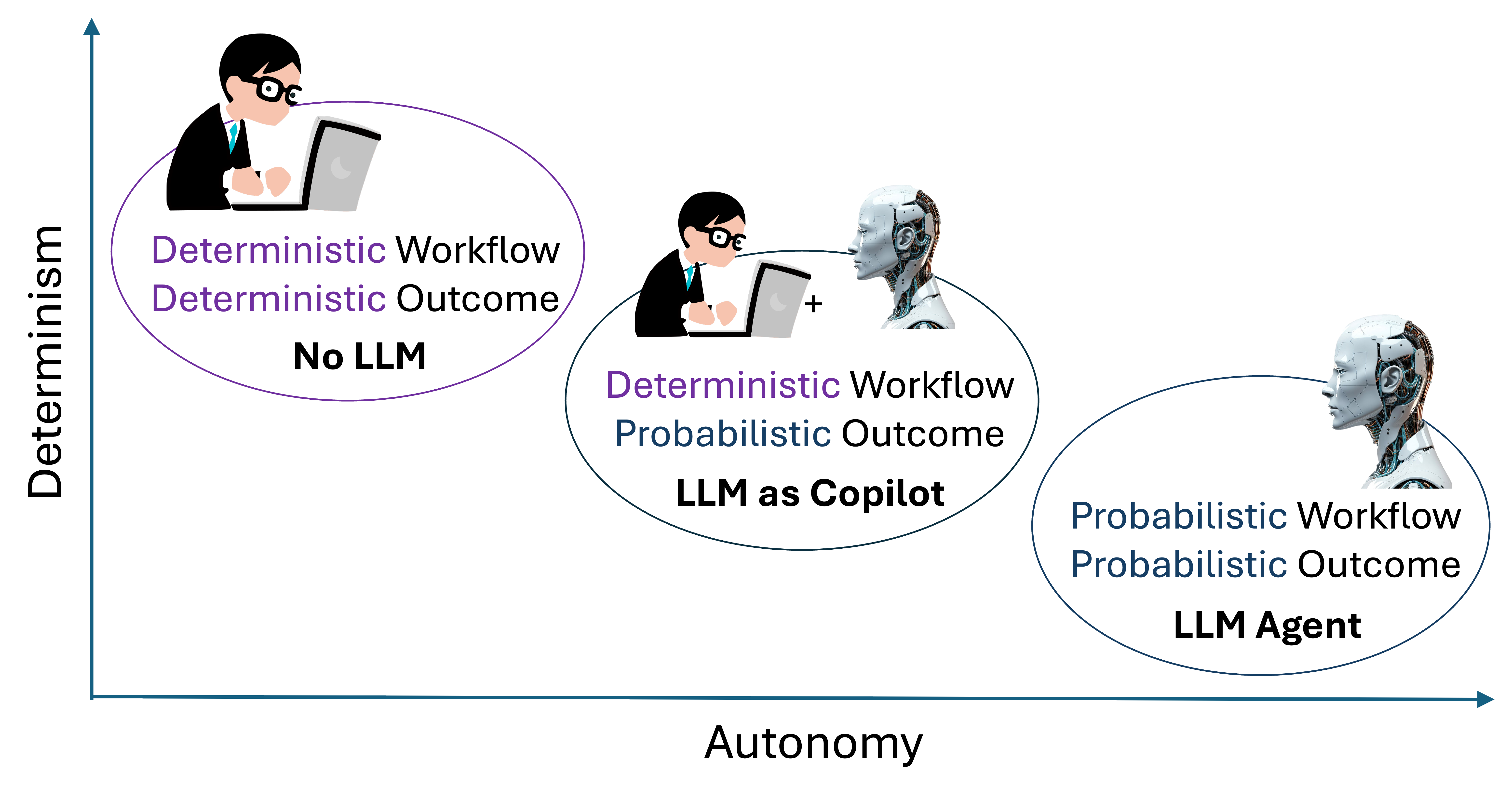}
    \caption{DAWN offers levels of determinism and autonomy required for different modes of agentic operation.}
    \label{fig:2}
\end{figure}

DAWN offers multiple operational modes application designers can choose from to orchestrate resources to accommodate tasks that require different level of determinism, as depicted in Figure \ref{fig:2}. Most new applications and solutions will likely integrate agents into existing business processes in copilot or autonomous mode as required.

\begin{itemize}
    \item \textbf{No-LLM Mode:} This mode is best suited for tasks that require high predictability and reliability, leveraging traditional algorithms and software tools to ensure consistent, repeatable results. Here, human operators manually design the workflow without using the Principal Agent and use only non-LLM resources and tools retrieved from the Gateway Agents to maximize determinism. In many ways, this is classical distributed software design and development.

    \item \textbf{Copilot Mode:} In this mode, agents act as intelligent assistants, enhancing deterministic workflows with creative input. Human operators design the workflow with the assistance of the Principal Agent's LLM-based planner and composer while using many of its constituent services and capabilities. Humans in this mode serve as overseers of the task planners and agentic workflow composers. Tools, agents, and agentic applications retrieved from the Gateway Agents are employed in the intervening steps to complete and execute the agentic operations. This mode is a hybrid between classical distributed software development and AI-based agentic software development.

    \item \textbf{LLM Agent Mode:} In this mode, agents operate autonomously, making decisions based on complex reasoning driven by prompting, LLM-as-a-judge \cite{zheng2023judging}, fine-tuning, etc. Here, the Principal Agent defines the workflow and retrieves the necessary resources from the Gateway Agents autonomously. This mode is particularly effective in for complex tasks where business logic is fuzzy and there is tolerance for stochastic results. Software developers play a significant role in the application design, workflow development, and testing phases for business and consumer applications that demand robustness and repeatability. They may play a lighter role for applications that are conversational in nature.
\end{itemize} 

DAWN implements safety, security, and compliance as a foundational layer within the architecture. It requires that all Gateway Agents meet minimum requirements across these three modes in order to participate. Principal Agents are always under the purview of the owning organization and hence all of the parent organization's regulations apply to its construction and management, while Gateway Agents may be under the purview of partners, vendors, and other third parties. This means its internals may be proprietary and not transparent, but its capabilities, access methods, and communication channels are subject to conditions and constraints. The DAWN framework has several key functions.

\begin{itemize}
    \item \textbf{Multi-agent collaboration:} DAWN builds agentic applications by facilitating collaboration among agents, enabling them to discover, communicate, and coordinate across different systems and organizations. Where more than one Gateway Agent is required to fulfill a set of tasks, the collected set of Gateway Agents facilitates the execution of complex tasks by harnessing the individual expertise of agents to collaborate effectively. DAWN leverages Gateway Agents to link agents globally. These Gateway Agents maintain agents themselves or link to registries that store agents' profiles and destinations. In the event agent registries have tagged agents for use with tasks or if agent profiles carry track records of prior use, the Gateway Agent uses that information for agent selection. It may also use its intrinsic capabilities to perform additional inspection of agents in response to requests from the Principal Agent.
    
    \item \textbf{Flexibility:} DAWN introduces a versatile approach that adapts to a wide range of task requirements, from fully deterministic processes to developer-driven design to dynamic decision-making. This adaptability allows for the seamless integration of traditional software tools, LLMs, AI agents, and multi-agent systems ensuring that the most appropriate method is used in each context.

    \item \textbf{Interoperability:} DAWN is designed for seamless integration with a broad spectrum of software tools, LLMs, and AI agents, by wrapping them as REST APIs in a unified format. It ensures interoperability across diverse ecosystems, which makes it adaptable to wider range of applications that require a blend of agentic processes and traditional software systems.

    \item \textbf{Connectivity}: DAWN uses an open protocol to interact and communicate between distributed components. The Principal Agent maintains a separate channel that allows new Gateway Agents to request connections. These connections are accepted after due vetting. Similarly, communication between the Principal Agent and Gateway Agents occurs over encrypted channels using HTTPS. A similar arrangement exists between Gateway Agents and all agentic resources. Gateway Agents belonging to private entities may define their own private protocols and encrypted transport mechanisms to communicate with agent registries. 

    \item \textbf{Comprehensive Security and Compliance:} The inclusion of a dedicated safety, security, and compliance layer underscores the architecture's commitment to maintaining high standards of data protection, user privacy, and regulatory compliance. This layer ensures that agentic operations within the framework adhere to stringent security protocols, protecting against unauthorized access and ensuring that the system meets industry-specific compliance requirements. Different organizations will demand varying levels of safety and security for handling agentic inputs and outputs and may specify these as a condition for participating in that organization's implementation of the DAWN architecture. These may be mandated by their own internal security, privacy, and compliance requirements, in additional to governmental regulations. 
\end{itemize}

\section{Related Works}

\subsection{LLM Agents}
The concept of agents has been a staple in the field of artificial intelligence for decades\cite{wooldridge1995intelligent}. However, it wasn't until LLMs were used as reasoning engines within these agents that the concept truly began to flourish\cite{wang2024survey}. An LLM agent typically comprises three key components: a reasoning engine, action space, and memory unit. The reasoning engine, powered by an LLM, is equipped with advanced reasoning strategies such as Chain-of-Thought\cite{wei2022chain}, ReAct\cite{yao2022react}, and Tree of Thought\cite{yao2024tree}, enabling it to process and respond to complex queries. The action space is defined by a set of tools that the agent can utilize to interact with its environment and obtain feedback, facilitating dynamic decision-making. A memory unit plays a critical role by storing the reasoning trajectories, conversation history, and relevant context, allowing the agent to accumulate and apply knowledge over time. As a result, hundreds of such agents are being developed daily, with the expectation that they will soon transform traditional software and human processes, leading to enhanced productivity, improved user experiences, and greater operational efficiency.

\subsection{Multi-agent Systems}
While individual agents have shown great promise, its true potential lies in the development of multi-agent systems. Human society has demonstrated that collaboration and teamwork often lead to superior outcomes, but the art of effective teamwork is complex and fraught with challenges\cite{salas2000teamwork}. Early efforts in building multi-agent systems have revealed tremendous potential\cite{chen2024agentverse, hong2024metagpt, qian2023chatdev, liu2023agentcoder, lin2023medagents, li2023camel}, but these works also highlight significant difficulties. Issues such as inefficient communication\cite{li2023camel}, increased computational costs, unreliable execution and communication sequences\cite{xia2024agentlessdemystifyingllmbasedsoftware}, and vulnerability to attacks\cite{ju2024floodingspreadmanipulatedknowledge} have underscored the need for a reliable platform to support agent collaboration. Additionally, scalability remains a critical concern: as the number of agents increases, introducing them to one another and ensuring that tasks are assigned to the most suitable agents becomes increasingly challenging. Addressing these concerns is essential for realizing the full potential of multi-agent systems in a wide range of applications. 

\begin{figure*}
    \centering
    \includegraphics[width=\textwidth]{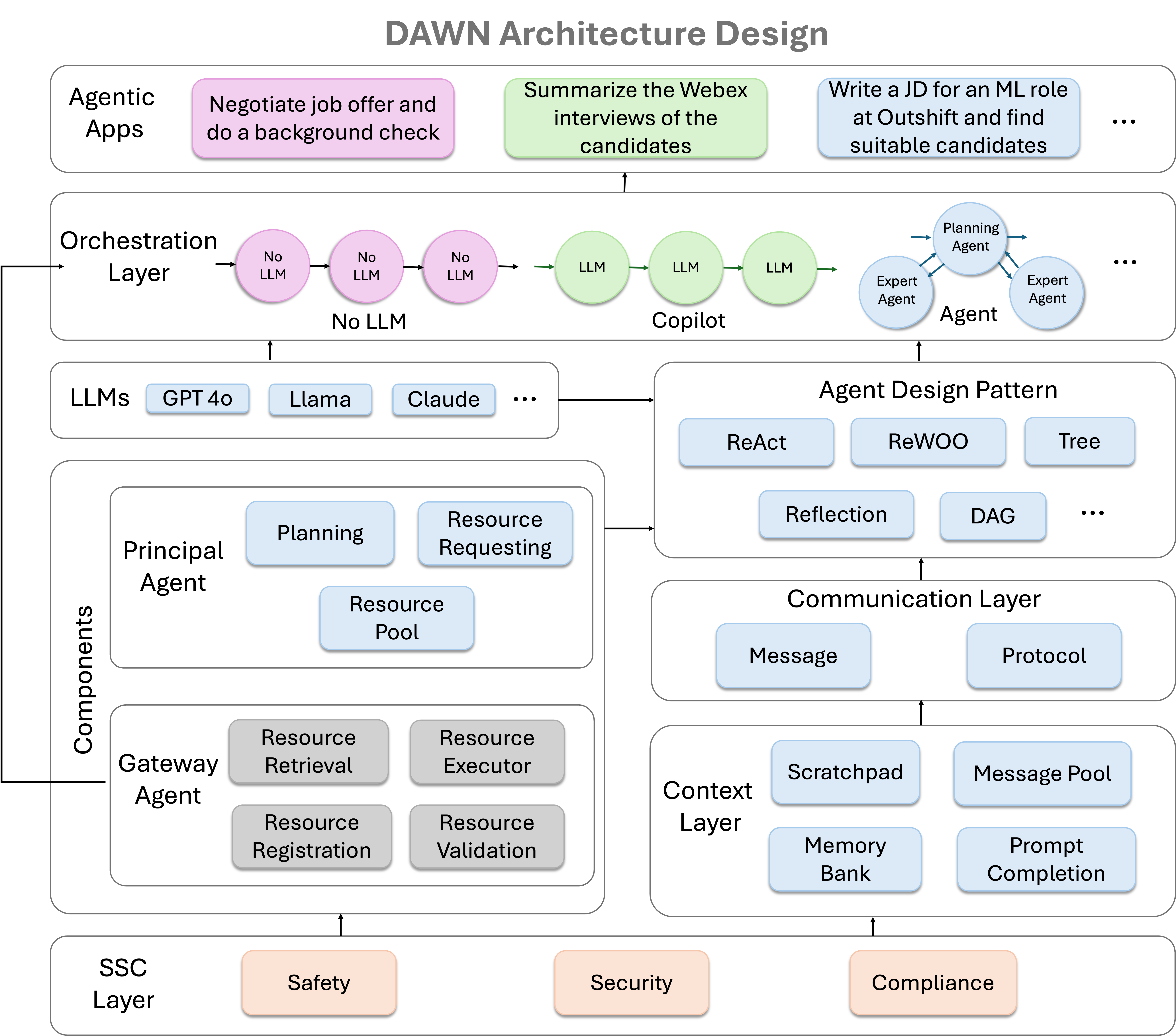}
    \caption{DAWN reference architecture}
    \label{fig:3}
\end{figure*}

\section{DAWN Architecture Detail}
The DAWN architecture is a modular framework designed to orchestrate collaboration of distributed agents. It supports agentic applications to be built with various operational modes, including No-LLM mode, Copilot mode, and Agent mode, to support tasks ranging from deterministic workflows (e.g., job offers and background checks) to creative and autonomous processes (e.g., interview summaries and job description compliance remediation). Central to the Agent mode is the Principal Agent, which plans tasks and requests corresponding resources (tools, LLMs, and agents), and multiple Gateway Agents, each of which independently oversees resource retrieval, validation, and execution required for the tasks identified by the Principal Agent. DAWN uses a robust communication layer to facilitate seamless interaction between the Principal Agent, Gateway Agents, and registered agents distributed around the globe. A context layer within Gateway Agents maintains memory and task history, ensuring that the requisite information is exchanged efficiently via the communication layer. The safety, security, and compliance (SSC) layer is responsible for the adherence of Principal and Gateway Agents to strict standards for the safe operation of agents during data and control interchange. Details of the reference architecture are shown in Fig. \ref{fig:3}.

Each organization may build and host one Principal Agent as the central portal for all of its users. From the organization's software developer’s perspective, a desktop client is installed locally and includes access to their Principal Agent with its context, orchestration layer, and SSC layers, and one or more Gateway Agents. Some of these Gateway Agents may be entirely internal to the organization while others may be external with customizations to fit the organization's particular needs. A developer may begin by selecting the appropriate operating mode for the application in question. If the user chooses the LLM Agent Mode, a chat-assistant interface is used to submit requests to the Principal Agent. If the application requested is a entirely new, the Principal Agent clarifies requirements with the user, interprets the request, devises a plan, and retrieves resources from Gateway Agents. (If a previously developed agentic workflow for that user task exists, then the Principal Agent presents that instead.) It then compiles an execution graph showing each subtask and corresponding resource, local or remote, which the user can approve. Upon approval, the orchestration layer oversees execution. Once execution completes, the user receives the final result. Because the Principal Agent runs locally in an organization's network (or equivalently, in that's organization virtual private cloud or VPC), user data resides in the local context layer, and transactions are overseen by the SSC layer, data privacy concerns are minimized.

\subsection{Principal Agent}
The Principal Agent acts as an autonomous planner-composer and central orchestrator within the DAWN framework. Its primary functions include creating a plan to execute customer requests, searching local resources, requesting additional resources from the Gateway Agents when necessary, assessing resources returned from Gateway Agents for best fit, and delegating subtasks' execution to the corresponding resources. Compared with the peer-to-peer communication pattern proposed by IoA\cite{chen2024ioa} and AutoGen\cite{wuautogen}, using the Principal Agent as the central planner and orchestrator ensures an orderly and structured workflow that is efficient, reproducible, and easier to trace and restart if prematurely stopped. The Principal Agent first plans, seeks resources that meet its plan, and composes the task sequence using the retrieved resources. Orchestration, or the execution of the agentic resources, then begins with the planning using the cached information gathered during the compose phase to implement the sequence of resources required to fulfill the user's intent.

The Principal Agent is the primary interface within any organization to develop, deploy, and operate agentic applications and resources. End users may use applications already developed for most of their daily needs or develop new ones on the fly by assembling available agents. Organizations use software applications and tools developed by large software houses to run their businesses. In the agentic era, these will be replaced or augmented by agentic applications living alongside classical ones. Agents that fulfill many new software functions will become available via Gateway Agents. Gateway Agents serve as subcontractors for fulfilling agentic needs. The separation from the Principal Agent provides for separation of duties and function, private ownership, intellectual property protection, and varied levels of security and compliance flavors. Gateway Agents are by definition headless in the DAWN architecture. They exist to provide specialized and custom services. 

The Principal Agent is a multi-tenant, multi-user application made of highly available constituent services that may be independently hosted and deployed across the organization, its private data centers, or on cloud data centers. In this regard, the Principal Agent is no different from an organization central web portal offering a host of services. The main difference is that the Principal Agent is also an autonomous AI-driven developer toolkit that can concurrently plan and compose multiple new applications and manage existing ones. And it has a built-in run-time environment to deploy and execute software. Each of its component services may be scaled independently. Principal Agents may be configured to reach out to internal and external Gateway Agents based on demand. Since most agents are hosted on or available via Gateway Agents, the availability of Gateway Agents serves as another scaling mechanism. Frameworks like Agntcy \cite{cisco_agntcy_2025} envision globally replicated and synchronized agent directories (registries that describe agent function and capability). Gateway Agents will have ready access to these directories and registered agents on-demand.

\textbf{Reasoning, Planning, and Adaptive Decision-Making:} The Principal Agent utilizes an LLM to understand the user's intent, develop a plan, and decompose it into manageable subtasks. This feature is evident when DAWN operates in copilot or agent modes. The Principal Agent's planner uses ReAct \cite{yao2022react}, as the default reasoning strategy, which follows a cycle of reasoning, acting, and observing to actively move through subtasks. This dynamic approach enables the Principal Agent to adjust to complex and dynamic environments. For example, an end-user might ask the Principal Agent to plan a trip. The Principal Agent would first interpret the goal end-to-end, and then create a plan with subtasks like booking a flight, reserving accommodation, and arranging local transport. If the flight turns out to be unavailable, the Principal Agent will search for alternative way such as train and bus, and modify the plan on the fly. The developer of the application decides how to prompt the planner component during development of this travel application. While ReAct works well in most use cases, developers may adopt other reasoning strategies. The Reasoning Without Observation (ReWOO) strategy \cite{xu2023rewoodecouplingreasoningobservations} is more suitable for less dynamic tasks where computational cost is a priority. ReWOO creates a comprehensive end-to-end plan without adjusting it during execution. HuggingGPT\cite{shen2024hugginggpt} uses this strategy to process multi-modal data by calling machine learning models available on HuggingFace\cite{huggingface} and achieves good performance with minimal LLM calls. Alternatively, the Tree-of-Thoughts (ToT) strategy \cite{yao2024tree} encourages the agent to explore multiple reasoning paths and self-evaluate them, offering more deliberate decision-making but at a higher computational cost. This approach is beneficial when performance is critical and cost is secondary. In Language Agent Tree Search (LATS)\cite{zhou2024languageagenttreesearch} and ToolLLM\cite{qin2024toolllm}, the authors combine the ToT strategy with other techniques to achieve the best success rate that surpasses ReAct. Developers may experiment with many of these before deciding on the best course that meets all of their application needs.

\textbf{Local Resource Pool:} The Principal Agent maintains a local resource pool that stores essential tools like calculators and web searchers, along with caching (references to) resources retrieved from Gateway Agents using a Least Recently Used (LRU) strategy. This setup enables the Principal Agent to efficiently handle simple or repetitive tasks without repeatedly consulting Gateway Agents, significantly reducing discovery and communication overhead while enhancing performance during task orchestration. The Principal Agent does not cache the agent itself if it is registered with a Gateway Agent -- only its location and capability are used by reference within the application workflow. Security and compliance are example services whose agents may live within the Principal Agent's orbit since they have a pivotal role to play in what information is sent to an LLM or received via chat response. Other resources that may exist locally include vector databases used in support of retrieval augmented generation (RAG) systems.

\textbf{Resource Requesting:} After developing a plan, the Principal Agent summarizes the plan and its context -- such as completed steps, conversation history, and user preferences -- into queries and sends them to all connected Gateway Agents in an effort to retrieve the most appropriate resources to execute the plan in its entirety. Gateway Agents may not being able to fulfill the entire plan because they don't have the resources to fulfill them all or because the resources required are down at the moment of request or because the task is niche and no agent exists for it. If there are any suitable local resources, then the Principal Agent will not seek send out request for those. With use, the Principal Agent's contextual memory component learns both the functional and non-functional capabilities of different Gateway Agents it interacts with. Over time therefore the Principal Agent may acquire the necessary statistics to prefer some Gateway Agents over others for a given resource or capability. At the very least this gathered data can advise the agentic developer in the selection of preferred Gateway Agents in the same way sub-contractors earn the trust of the primary contractors in manufacturing or supply chains. 

\textbf{Resource Execution}: Gateway Agents are expected to apply their programmed discretion in searching, matching, and retrieving the most applicable resources. When requests are sent to multiple Gateway Agents, it is only natural that duplicate resources or resources with overlapping capabilities are returned to the Principal Agent. The Principal Agent is responsible for analyzing the returned resources, assessing the quality of each resource, determining which agentic resource or resources may be applied to each task, and then retaining just those resources. During the execution of the application workflow, the run-time component of the Principal Agent begins execution of the task graph. The execution continues until it hits a node where a new search for resources is required to complete the remaining subtasks. (This situation can arise if during the compose phase of the process, the returned resources from the Gateway Agents only partially fulfill the customer need.) The planner then reaches out to the Gateway Agents again with the remaining requirement. This process repeats until the user's request is fully addressed. (In some cases, the response may be null and therefore some organizations may be better served if only fully qualified software applications with available agents are released to users.) It should be noted that Gateway Agents return references to agentic resources. For instance, some agents may actually execute on their preferred cloud service provider locations. In other cases, the agents may be imported into the Gateway Agent and rely on run-time environment of their host Gateway Agent. In both cases, only callable references to resources are passed between agents, Gateway Agents, and Principal Agents. And so, when the Principal Agents executes resources from a Gateway Agent, it is in reality making an API call to the resource and passing the payload to the concerned Gateway Agent via JSON or similarly packaged property file format.

In many agentic implementations, the action space or tool-set is limited to the local resource repository. This is typically due to three factors: 1) specialized agents naturally operate within a limited action space\cite{ALFWorld20, zhang2024proagent, abuelsaad2024agenteautonomouswebnavigation}, 2) the context window of LLMs restricts the number of resources they can handle, and 3) if the agent has more than 20 tools to choose from, the accuracy in selecting the right tool drops as suggested by OpenAI\cite{openaifunctioncalling}. (During experimentation the authors found this number to be less than 10.) The Model Context Protocol, MCP \cite{modelcontextprotocol_intro_2025}, was invented to address this shortcoming. However, fully autonomous agents need to navigate real-world problems where the action space is vast, and numerous resources are available. DAWN's rationale to use distributed Gateway Agents follows directly from this tenet.

For example, if a Principal Agent needs to perform an action beyond an LLM's capabilities -- such as online search, booking a ticket, or writing code -- it will request the relevant resource (e.g., a search tool, booking agent, or coding agent) from Gateway Agents. This flexibility enhances the Principal Agent's autonomy and allows it to handle a broader range of use cases. However, flexibility of this type comes with a cost. The Principal agent might get overlapping, conflicting, malicious or faulty resources. Overlapping or duplicate resources require special handling to determine which resource is best in terms of functional as well as non-functional performance characteristics. In case malicious agents are introduced, whether inadvertently or otherwise, these agentic resources will render the entire distributed agentic application vulnerable to attack and worse, loss of faith. This is precisely why the safety and security of such applications are of crucial importance. Such faulty resources are an inherent part of the internet as well and people have been working to keep the defense line strong. This is the price we must pay for an open, participatory system. The alternate is a closed one and the trust that one provider will make it safe, secure, but inflexible. 

\subsection{Gateway Agent}
The Gateway Agent plays an essential role in connecting globally deployed resources (tools, agents, and agentic applications) with the Principal Agent. The Gateway Agent offers several services of which two are primary: a registry where resources are registered and an intelligent match-and-retrieve service for searching relevant resources given a query or task list. When the Principal Agent sends queries describing its need in natural language to the Gateway Agents, the Gateway Agents in turn search their registries and return the most appropriate resources to the Principal Agent. Each Gateway Agent may apply various filters and guardrails to select and trim the number of resources returned. Security measures such as access control and resource testing are integrated to reinforce safety of the system.

The Gateway Agent uses an open protocol, promoting decentralization and collaboration. Allowing different organizations and individuals to create and manage their own Gateway Agents leads to a global ecosystem of interconnected resources, encouraging innovation and diversity in problem-solving. It also adds a level of scalability and flexibility, as users would have access to a broader range of capabilities and knowledge through international cooperation. The key functions of Gateway Agents are listed below:

\textbf{Resource Registration:} The Gateway Agent hosts and maintains a registry where developers can register their resources (tools, agents, and agentic applications). A successfully registered resource will be accompanied with crucial details, such as the resource's name, application programming interface (API), input and output schema, and API documentation. Information such as resource description, usage examples shall be included so that LLMs can understand what the resources are and how to use them through in-context learning. Gateway Agents may be equipped to run agentic resources within its confines or expect that agents have their own run-times.

When a resource is first registered with a Gateway Agent, the Gateway Agent has only limited information about it. It is screened for safety, security, compliance and other readily visible and externalized characteristics. Over time, the Gateway Agent gathers utilitarian characteristics such as P50/P90 latency, cost, completion rate, prior use, etc. The registry may also include metrics such as success rate, average execution time, availability, and load capacity. These metrics allow the orchestrator to optimize resource utilization and then use it to score, rate and select the best resource for a given user function.
\begin{figure*}
    \centering
    \includegraphics[width=\textwidth]{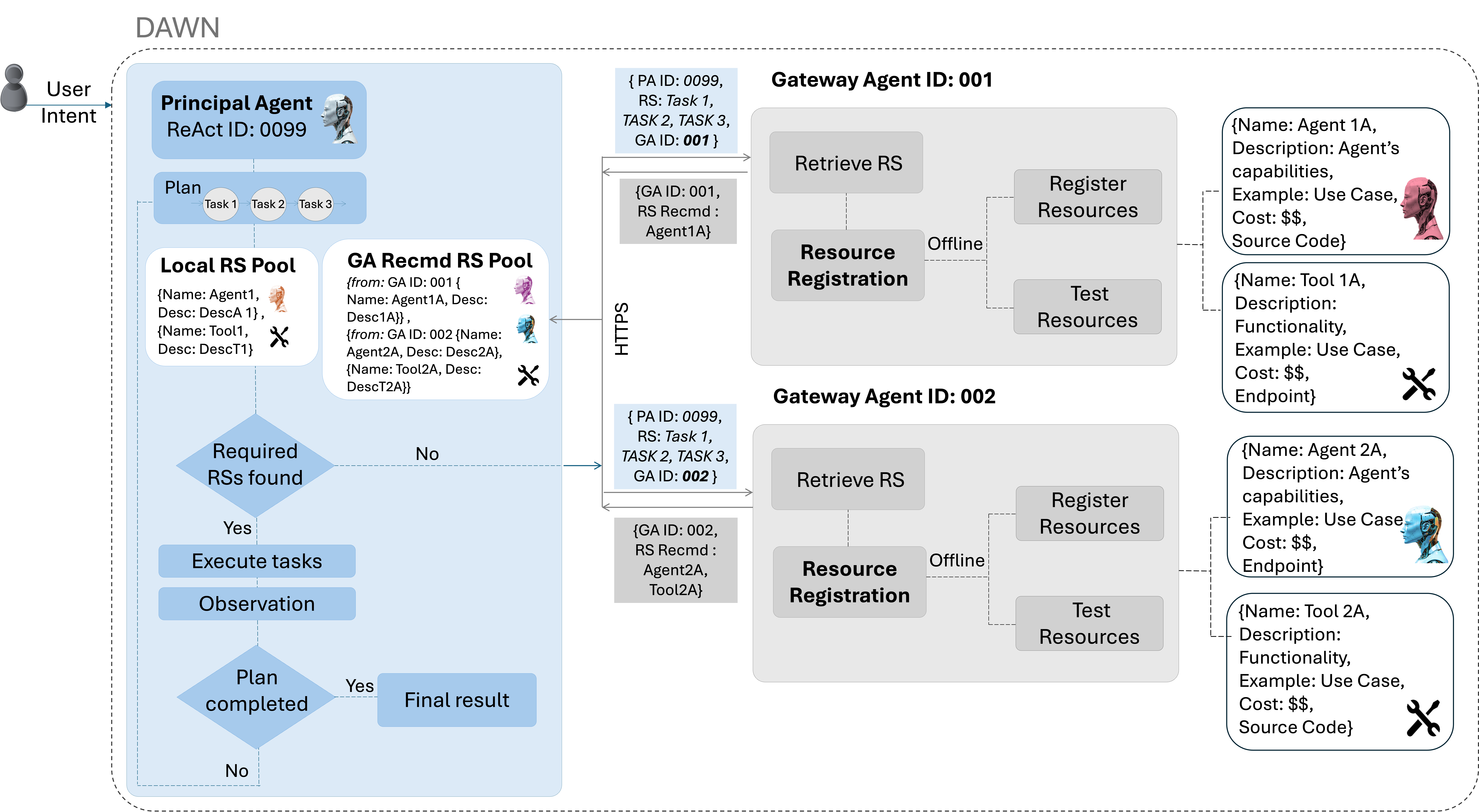}
    \caption{Principal Agent and Gateway Agents workflow. When invoked with an intent for an application or solution, the Principal Agent first plans, produces a task list, and then requests resources from the Gateway Agents. The Gateway Agents return the most suitable resources to the Principal Agent, which is then used to first compose, then orchestrate, and next, execute the task list.  }
    \label{fig:4}
\end{figure*}

\textbf{Resource Executor:} For agents developed in different ecosystems to effectively communicate with the Principal Agent, execute tasks, and deliver results, they must adhere to a common protocol. In this case, the use of natural language specifications for agents and API descriptions can serve as the established common protocol. This protocol ensures seamless interaction and interoperability. As developers register their resources on the Gateway Agent, they are required to provide descriptions of their resources. These descriptions include details about the capabilities of the resources and instructions on how to utilize them. 

LLMs can automatically communicate with RESTful APIs to execute tasks through them. RestGPT\cite{kim2024leveraging}, for example, showcases this potential by leveraging the coding capabilities of LLMs to generate parsing code based on structured schemas. This approach automates both the generation of API calls and the parsing of responses and furthermore the LLM acts as a backup parser when necessary. 

When incorporating agents from other frameworks and utilizing their manifest files, the Gateway Agent sends the manifest files to the Principal Agent. This enables the Principal Agent to communicate through a valid schema and ensures that the LLM can parse the response accurately based on the agent's manifest file. By utilizing this mechanism, effective communication and parsing are maintained between the Principal Agent and other agents from different frameworks and allow for interoperability of such network of agents. 

\textbf{Resource Testing:} Testing resources is crucial to ensure their proper functioning. Developers and providers who own the runtime environments are responsible for periodically testing and validating their resources. In addition to functionality testing, resources should be tested for connectivity, performance, security, and other factors \cite{rapidapi_api_testing}. For instance, it is important to evaluate how well a resource handles different levels of traffic. Various commercially available tools like Apache JMeter \cite{apache_jmeter}, Gatling \cite{gatling_tool}, and Postman \cite{postman} provide the capability to test resources with a high volume of concurrent users. 

If a resource provider fails to provide validation, the Gateway Agent will temporarily suspend the resource until the required tests are successfully completed, and the validation is provided. Likewise, resources managed by the Gateway Agents will undergo a policy check and alignment with guidelines through empirical testing and automated review. If a resource fails to pass this validation process, indicating discrepancies between its description and actual behavior, the Gateway Agents will disqualify the agent from participating until a more accurate description is generated. 

\textbf{Resource Retrieval:} A Gateway Agent handles queries from the Principal Agent, and searches its registry for the most relevant resources. Similar to traditional information retrieval methods \cite{manning2009introduction}, resource descriptions and examples may be treated like documents. Semantic search \cite{stoica2003semantic} may be used to identify the most relevant resources by converting both the queries and resource descriptions into vector embeddings and then performing similarity searches based on those embeddings. To further improve search accuracy, additional techniques like attribute filtering and keyword search are incorporated. Moreover, fine-tuning LLMs on query-resource pairs is another viable approach to enhance search results \cite{patil2024gorilla, qin2024toolllm}. While DAWN imposes some base functional and non-functional requirements on Gateway Agents along with stringent safety and security guardrails, its internals are left to the discretion of the Gateway Agent's owner. Some Gateway Agents, however, may only have traditional software and classical data repositories.

\textbf{Multiple Gateway Agents:} Organizations and developers around the globe will build and manage their own Gateway Agents. While DAWN prescribes the components and services included in a Gateway Agent, the implementation is left to the organizations that build and manage them. To participate in a agentic solution with a Principal Agent, the Gateway Agent must have an open protocol. Each Gateway Agent offers unique resources proprietary to the entity in question (organization and its developers). Gateway Agents are also by design hybrid entities combining classical and agentic components to cover both AI and non-AI functionality. At any given time, the Principal Agent has a list of Gateway Agents it is connected with. These are loosely assembled entities along the lines of cloud-backed APIs. The Principal Agent submits tasks and user intents to the connected Gateway Agents which then return the best resources available that may address all or a subset of the tasks or subtasks associated with the user request. The Principal Agent collects the results and selects the highest-ranked (and most appropriate for the task) resources from them. This selection is based on factors like relevance, latency, cost, reputation, prior use, success and failure rates, and other metadata collected by the Gateway Agents about these agents over time. Gateway Agents retain and learn similar data about agents attached to themselves and also which services are requested and used by Principal Agents. This metadata may include information such as rating, latency, and performance of the registered agents. If the returned resources serve to complete the user tasks, then the Principal Agent's task list is complete. If not, the cycle continues as directed by the Principal Agent.

Gateway Agents may disconnect for maintenance or upgrades when idle or if its pending requests can be queued. Gateway Agents designed with memory and context mechanism to hold unaddressed user requests in a queue are able to connect and disconnect on their own schedules. A new Gateway Agent may join the Principal Agent's list at any time but new Gateway Agents may only participate in a subsequent user request from the Principal Agent. This stands to reason as it might not have the context to address an ongoing task list.  The Principal Agent continuously evaluates Gateway Agents based on the quality of the resources they provide. This evaluation helps assign weights and ratings to Gateway Agents and the resources they offer. Furthermore, end-users and developers have the option to rate and assess the final execution of their request, including how each element was handled. These user ratings may also impact the ratings of Gateway Agents. The Principal Agent also uses its own memory module to learn from its own experiences. Fig. \ref{fig:4} illustrates the workflow involving a Principal Agent and two Gateway Agents, incorporating most of the functionalities discussed above.

\subsection{Orchestration Layer}
The Orchestration Layer enables the system’s versatility by supporting different operational modes, including No-LLM, Copilot, and LLM agent modes. These services enable enable multi-tenant and multi-user concurrency and keep each operating application distinct and separate from all the others. When one such application is composed by the human user or the Principal agent, the Orchestration Layer services are responsible for enforcing the rules, logic, and sequencing necessary to ensure that each subtask is executed in the correct order. Internally, the task graph created from the each task list is implemented as a finite state machine whose nodes are agents or agent ensembles. Additionally, the Orchestration Layer maintains the global state, tracking the progress of task execution across all nodes in the workflow and across all the participating Gateway Agents (and global agentic resources). This global state ensures that each resource has the contextual information needed to perform its assigned subtask effectively. By maintaining and updating this global state, the system ensures that subtasks are executed with full awareness of prior actions and the broader context of the overall task.

\subsection{Communication Layer}
The Communication Layer is responsible for managing the flow of information between various system components and resources, ensuring seamless interaction across the platform. Its inherent services handle the protocols that ensure messages are properly parsed, interpreted, and routed to their correct destinations. Two primary types of messages are exchanged among the system components and resources.

The first type are the messages exchanged between the Principal Agent and its connected Gateway Agents. The payload from the Principal Agent contains the subtasks and their descriptions. This may involve data cleansing and purposeful query generation that focuses on the specific information or task needed, without providing excessive user details. During task execution, if a resource fails, the Principal Agent will reach out to the Gateway Agents to either repeat instructions or gather logs and metrics. The Principal Agent will provide the current subtask and its description as well as the completed subtasks and interaction history. This additional context enables Gateway Agents to make informed decisions about which new resources to retrieve. In response, the Gateway Agent then returns a list of suitable resources along with manifest files that contain the full record of the functional and non-functional capabilities of the resources. All of these interactions happen over channels set up by the Communication Layer.

The second type of message is the execution command sent by the Principal Agent to the Gateway Agents' resource executors. These messages contain the necessary execution details, such as the resource endpoint and input parameters required for the task. A given resource executor may then respond with the result of the execution. In the event of an error during execution, the message will include an error notification to inform the Principal Agent, allowing for appropriate error handling and adjustments.
 
\subsection{Context Layer}
The Context Layer is vital for enabling agentic operations by managing memory and task-related context. It provides three key components for memory management: a scratchpad to track the agent's internal reasoning process, a message pool to record communications between the Principal Agent and Gateway Agents, and a memory bank to store long-term data and user preferences, supporting long context, in-session, cross-session, multi-turn conversations resulting in personalized user experience. The Context Layer services dynamically assemble prompts using data from both the message pool and memory bank, ensuring that the Principal Agent is aware of the overall task, task progress, and available resources for all tasks and users that may be using the system at any given time.

\begin{figure}
    \centering
    \includegraphics[width=1\linewidth]{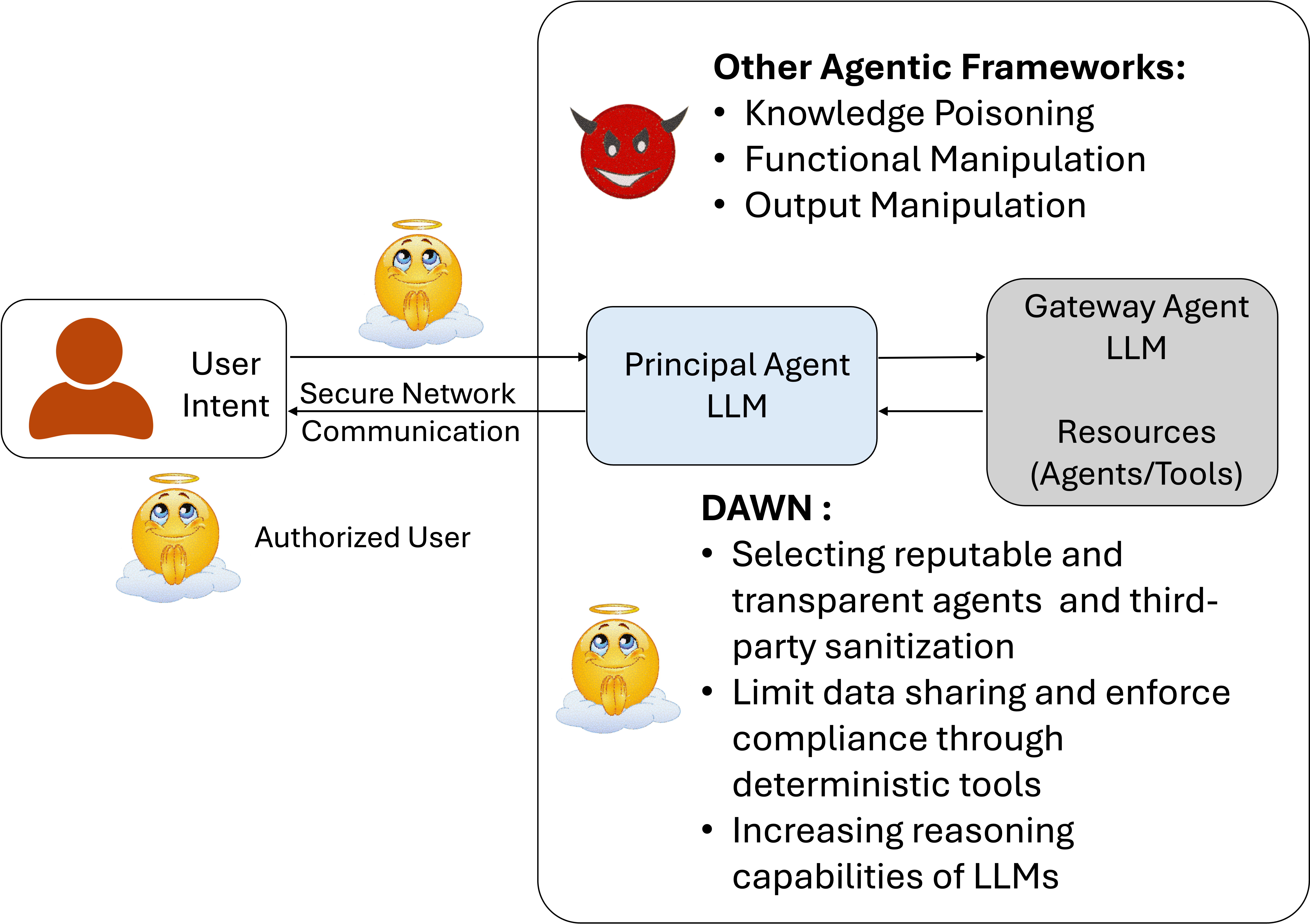}
    \caption{Possible attacks on agentic networks and how DAWN mitigates such attacks. The Principal Agent and each of the Gateway Agents implement safety, security, and compliance screening to scrub agentic resources' inputs and outputs.}
    \label{fig:5}
\end{figure}

\subsection{Security, Safety, and Compliance Layer}
The Security, Safety, and Compliance Layer establishes a robust security foundation for the entire framework, addressing the safety, security, and compliance needs of the system. All frameworks and platforms that employ distributed resources must implement robust security systems to avoid risks that may be introduced when unsecured resources are connected to an enterprise's gateway agent. While extensive research and practices have been published on traditional common security issues such as authentication, access control, privacy, and policy enforcement \cite{granjal2010secure , jebri2015efficient, jing2014security}, an agentic network equipped with LLMs introduces additional security and privacy threats especially as resources worldwide begin collaborating \cite{owasp2025llm, shavit2023practices}. The DAWN framework leverages available safety and security components and services to eliminate network attacks through traditional measures. Additionally, the outlined steps provide multiple layers of security within the agentic network, ensuring the system’s integrity, trustworthiness, and regulatory compliance. For LLM agents, two primary sources of threats exist.

First, there are vulnerabilities inherent to LLMs, such as hallucinations \cite{huang2023survey}, attacks that leverage tuned instructional methods like jailbreaking \cite{li2023multistep, shen2024donowcharacterizingevaluating}, and prompt injection \cite{10.1145/3605764.3623985}. Second, there are threats specific to agents collaborating in an agentic network like DAWN. A notable example is knowledge poisoning, where malicious data is integrated into the LLM’s knowledge base. PoisonedRAG \cite{zou2024poisonedragknowledgecorruptionattacks} exemplifies this type of attack, where the corrupted knowledge base causes an agent to generate attacker-chosen responses to targeted questions. Another attack vector is functional manipulation. In the LLM agent’s workflow, after the Principal Agent takes an action, it processes the results and moves to the next step. An attacker can manipulate this process by inserting harmful prompts, thus directing the agent to perform unsafe tasks \cite{zhan-etal-2024-injecagent}. For instance, in a recruiting scenario the Principal Agent might retrieve a fraudulent review written by an attacker to recommend hiring an unsuitable candidate for a job. Agents are vulnerable to such unauthorized actions and similar manipulations.

There is extensive research dedicated to addressing both the inherent vulnerabilities of LLMs and attacks on agentic networks\cite{he2024emerged} . To protect the DAWN framework against the knowledge poisoning, selecting reputable and transparent resources is essential. These resources must provide clear data usage policies and strong security records \cite{zhang2024s}. Tools and agents should also undergo third-party testing and sanitization, as demonstrated by ToolEmu \cite{ruan2024identifying}, which proposes a framework for emulating tool execution through a language model. This approach allows scalable testing of the resources across various tool-sets. Defense against functional manipulation begins with proactive security measures. For example, when employing third-party LLM agents, users should limit data sharing, particularly of sensitive personal information (PII). DAWN supports dynamic configurations, particularly the co-pilot mode, that integrates LLM-based and deterministic workflows. This ensures privacy and enterprise compliance that are enforced by guardrails through algorithms designed to filter inputs and outputs\cite{welbl2021challengesdetoxifyinglanguagemodels,gehman2020realtoxicitypromptsevaluatingneuraltoxic}. Incorporating deterministic steps into the workflow allows the enforcement of organizational policies and ensures user compliance through a LookUp tool, addressing potential instances where LLMs may bypass established guidelines. Additionally, increasing the reasoning capabilities of LLM agents is essential to counteract output manipulation. In the DAWN framework, the impact of output manipulation attacks is reduced by augmenting the Principal Agent with reasoning capabilities and access to a memory layer. Additionally, the modular architecture design enables the tracing of agentic attacks. Fig. \ref{fig:5} shows the attacks specific to agentic networks and DAWN's approach in mitigating them.

\begin{figure*}
    \centering
    \includegraphics[width=\textwidth]{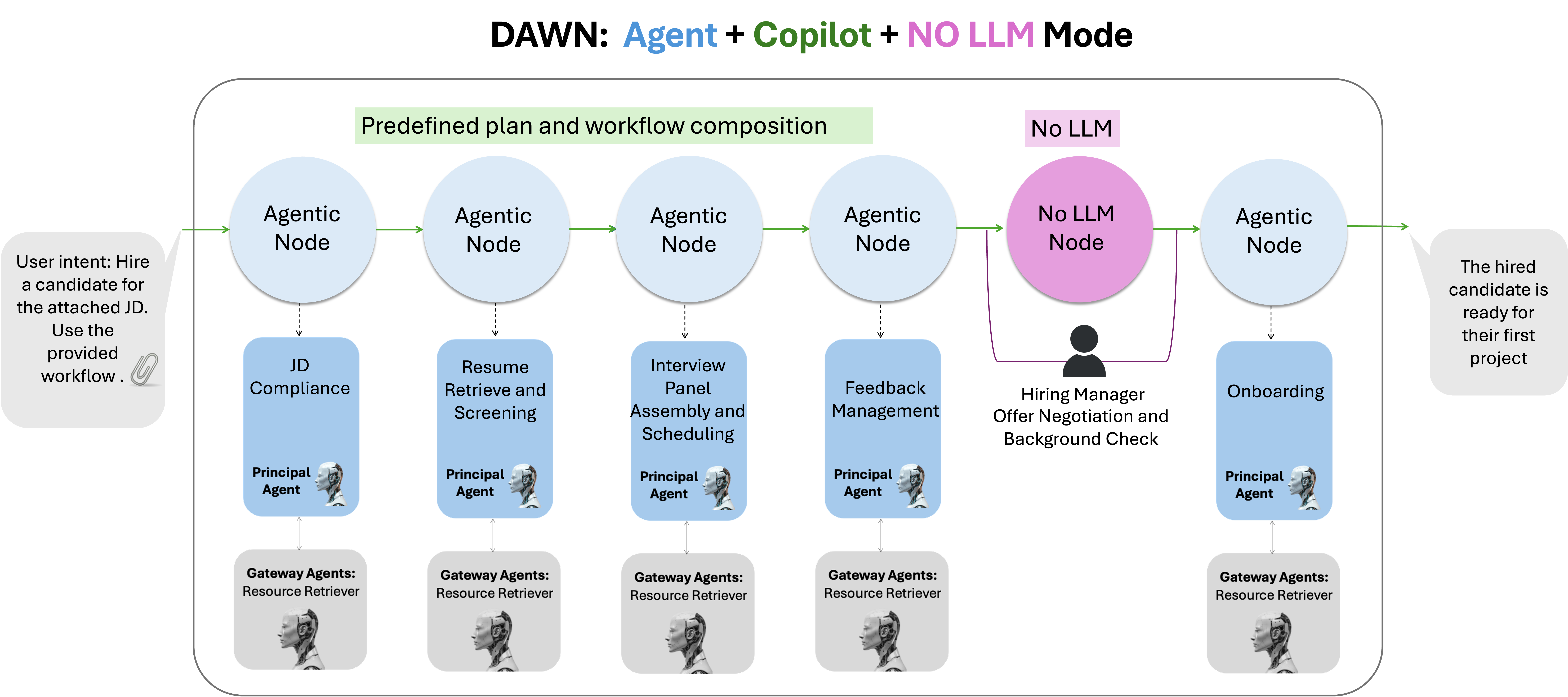}
    \caption{Application of DAWN in an Human Resources recruiting use case in Agent, Copilot, and No-LLM modes }
    \label{fig:6}
\end{figure*}

\section{Application of DAWN}
The modular design of DAWN framework provides users with the flexibility to select the most appropriate working mode for their application, tailored to the specific needs of various use cases. As outlined in the introduction of this paper, the primary modes include No-LLM, Copilot, and LLM Agent mode. For scenarios requiring high determinism, users manually define plans, select suitable resources through the Gateway Agent, and assemble a No-LLM application within the orchestration layer, ensuring precise and predictable outcomes. In situations where uncertainty is present and decisions must be made dynamically based on real-time feedback, users manually construct a decision graph, incorporating an LLM to intelligently select which subgraph to execute based on evolving conditions. For cases with flexible workflows, users delegate the entire task to the Principal Agent, which autonomously, or along with human oversight, plans, finds resources from Gateway Agents, and constructs agentic applications. This adaptability ensures that the framework can efficiently handle a wide range of complex and diverse tasks.

To demonstrate DAWN’s capability, a walk-through of the implementation of a Human Resources (HR) recruiting use case is presented below. Finding the right candidate for a specific role in any organization involves multiple steps that span weeks. Many of the steps are repetitive and therefore well-suited for agentic automation. They also require supervision and oversight at every stage. Most organizations have hiring processes in place and agentic software, such as this HR implementation of DAWN, needs to be built jointly with HR experts and according to that organization's policies and guidelines. Certain steps, such as offer negotiation with a candidate, require high security and sensitivity and are best handled by humans. Therefore, developers and users are advised to apply DAWN in copilot mode. 

Fig. \ref{fig:6} illustrates an HR example, a composite of Agent, Copilot, and No-LLM modes. This approach allows the Principal Agent to devise a plan once it is provided the requirements as instructions. The human then ensures that none of the essential steps in the hiring process is neglected. The supervision also ascertains that all of outputs that demand inspection are closely monitored. The workflow is a directed graph with most of the steps assigned to agentic nodes. These nodes leverage the autonomous agent mode where possible with the Principal Agent requesting resources from Gateway Agents. In fully autonomous mode, multiple specialized agents will be seamlessly coordinated by the Principal Agent’s orchestration and the Gateway Agent’s resource retrieval. However, here the deployment was supervised by an human. The agents and process flow deployed in this use cases were designed using the DAWN architecture and implemented using agents created using the LangGraph framework. Planning and composing was done using a Principal Agent and the agents themselves were registered with two different internal Gateway Agents. (While two were not strictly necessary, they illustrate how specialized agents may be registered with Gateway Agents with different owners.) The details of these agents are as follows.

\begin{enumerate}

 \item The hiring process begins with creating a job description or JD. JDs have specific characteristics, mandatory sections, and descriptions. For this implementation, the team developed an agent to write the JD for a machine learning engineer using descriptions gathered from the web. The agent was provided with sample JDs. This step was developed in agentic mode with developers (proxies for HR experts in this case) reviewing the output and iteratively modifying the prompts until satisfied. The DAWN framework was used for iterative development and testing, and its intrinsic SSC layer was used to make sure the output JD was compliant to company, federal, and state policies and regulations. JDs are short documents and take only a few seconds of LLM time to process and produce new versions.

 \item With a JD in place, the next agent in the workflow finds matching candidate profiles in LinkedIn using the LinkedIn API. The agent first pulls apart the different sections of the JD such as title, experience, education, skills, location, etc. It then applies these requirements to the LinkedIn advanced search API. Next, the LinkedIn filtering APIs are used to refine the results. Then, as a final step, the agent applies any organizational screens to further distill the results. This search-and-filter step is iterative and requires humans to review the number and quality of the retrieved profiles to adjust the agent's input constraints. The agent functions autonomously but its output needed to be adjusted based on the job description and level of the intended hire. Searching and filtering takes but a few seconds but altogether this step can take hours because of the inspection required.

 \item The third agent constructs the interview panel for each candidate for each job role. It uses the same pool of available interviewers and accesses their calendars to find empty slots. Scheduling itself is simple but more realistically contacting the candidate for available time slots is done by recruiters. In this implementation, only the calendars of the interviewers were checked. This agent interfaces with Microsoft Outlook to book these slots. No allowance was made for change requests though realistically that should also be accounted for. This agentic step was fully automated and took milliseconds to execute.

 \item While agents can listen in to interviews, transcribe, make summaries, and conclusions, that is often impractical, socially unacceptable, and in some jurisdictions, outright illegal. In this implementation, the agent sends reminders to interviewers to document and upload the interview feedback within a time window, gathers and aggregates all the feedback from the database, and compiles that into a single document in preparation for the next phase - the interview debrief and decision meeting. 

 \item Once a decision is made, all the materials required for the hiring manager to make the final decision are done using existing tools and processes. This is the No-LLM "agent" depicted in the figure. A lot of hiring experience and support from recruitment experts and financial advisors go into this step and it makes little sense to try and use AI except to gather information and make sure decisions adhere to policies and guidelines.

 \item The last agent sets up onboarding education for the new employee. It compiles the briefing package that includes reading material, self-training, people to meet, orientation meetings to attend, and other information required of all newbies. And then are special sessions for those who may have specialized responsibilities. In this implementation this was conceived as a list to complete with appropriate calendaring. The agent sends reminders and calendar events after building a customized onboarding and orientation program for the new employee based on level and specialized function.

\end{enumerate}

To measure the effectiveness of this DAWN architecture requires looking at success and failure rates in each component and each step of the workflow. Agentic software development covers the following agentic actions: plan, compose, discover, connect, orchestrate, observe, safeguard, and communicate. Therefore, each of these may be measured and evaluated, and taken together indicate how effective DAWN is. Of these, plan, and compose lie entirely within the Principal Agent's purview. The remaining functions are distributed between the Principal and Gateway Agents. This paper has already addressed connect, orchestrate, observe, safeguard, and communicate to some extent. For the most part these functions mirror those used in traditional distributed systems and may be evaluated along the same lines. Plan and compose are novel in agentic software development. Therefore, in this experiment only the plan and compose elements of the process was evaluated. Planning involves producing a task list from the application intent while composing involves producing a task graph (finite state machine) from the task list and finding and selecting the most appropriate agents to perform the actions. The following algorithm was used to measure the success rate of the planner:

\begin{enumerate}

 \item Provide the LLM with a fully qualified application intent and it outputs a task list. An LLM is prompted to generate 10 use case descriptions that demand the use of agentic AI, each with a task list and a graph of "tasks that are not just sequential". It produced 10 task lists and 10 task graphs. This serves as the ground truth. 
 \item Select a use case at random from the list and send it to the same LLM (as in step (1)) using the same prompt. Compare the task list against the ground truth. 
 \item Repeat step (2) 5 times. Measure repeatability in terms of all the tasks and the correct order of the tasks. (In some cases. reordering tasks may not matter and in these cases reordered tasks are marked correct provided they are complete in other respects.)
 \item Optionally, repeat steps (2) and (3) with other LLMs
 \item Compute the success rate.

\end{enumerate}

The 10 LLM generated use cases each had 6-10 tasks (task descriptions each 1-2 sentences), some that required a sequential application of agents, and others, more complex graphs. The computation for steps (2) and (3) above was done manually and using a different LLM as a judge. The success rate for task list creation ranged from 70-85\% -- the lower number when a different LLM (OpenAI 4o-mini) was used for assessment from the original one used for task list creation. The complexity of the tasks varied quite a bit across the 10 use cases and that might also explain some of the variance in the results.

Composition requires finding the best matching agents for a given task. This is modeled as an information retrieval (IR) task. The IR task was setup as a separate experiment to measure task-to-agent matching accuracy against a well-known public IR dataset with 200 agents and over 20,000 queries. Given the agent registry with agent names, descriptions, and other metadata, a two-stage information retrieval system was developed to find the best fitting agent for each task. Stage 1 is a retriever backed by a vector database and stage 2 is an LLM-based re-ranker. The retriever produces a fixed list of top-5 agents for each task in the dataset and the re-ranker uses those candidates and any additional contextual information provided to generate the best candidate/s. Two metrics, Normalized Discounted Cumulative Gain (NDCG, measured at top k) and Recall (Recall@k), were computed. After stage 2, NDCG at k=1 and k=3 was 0.71 and 0.78, respectively, while Recall at k=1 and k=3 was 0.71 and 0.82, respectively. 

Each of these agentic steps executed in seconds with most of the elapsed time spent in LLM interactions. When run end-to-end, the runs varied from eight to 20 seconds using GPT-4o with all of the variance in elapsed time attributable to LLM inference times. Every step of the hiring process allocated to an agentic node involved the Principal Agent and Gateway Agent working together, showcasing the copilot and autonomous capabilities of this DAWN implementation. In this HR application, the task graph (agentic workflow) awaits human interaction after each agentic step - a requirement of the designer. Once completed the workflow progresses to the next step automatically. The No-LLM workflow step demonstrates the inclusion of a traditional non-AI business process where the hiring manager takes over. This step includes salary estimation, negotiations, and candidate background checks which involve interactions between multiple parties and highly confidential and sensitive data that must be kept secure. Once complete, the workflow resumes where it left off, but requires specific prompting. (It also handles concurrency when multiple such requests are in flight, and prompting for resumption requires additional data to pick up the pending actions.) The No-LLM mode illustrates how the DAWN architecture allows for the integration of human-in-the-loop and determinism within an application's agentic workflow when necessary. This HR use case was implemented with the listed resources to demonstrate the functionality of this framework, and leverages both expert knowledge and the autonomy of using LLM agents in applications requiring sophisticated task execution.

For each step of the workflow, the Principal Agent communicated back and forth with the connected Gateway Agents for further action, ensuring the plan is regularly evaluated and the user interface (and user) is updated with operational status. This requires that state is continuously exchanged between the Principal and the Gateway Agents for error management, telemetry data collection, and status updates. This continuous communication is essential to support iterative and traceable problem-solving, and ensure transparency and accountability throughout the entire workflow execution process, thus making complex task management more effective. Yet another implementation using the DAWN architecture is described in this Cisco Tech Blog \cite{patel2025building}. This implementation also deploys one Principal Agent and two Gateway Agents but this time using three different agentic frameworks, LangGraph, Swarm, and AutoGen (or AG2), for agent creation, deployment, and execution. Execution times tracked what was measured in the HR use case. Together these two implementations demonstrate the flexibility and viability of the DAWN architecture.

\section{Conclusion and Future Direction}
The DAWN platform revolutionizes agentic applications by establishing a robust foundation for agent collaboration and communication. This framework offers a comprehensive strategy for integrating LLM-based agents into diverse applications, both new and existing. Its modular design, featuring key services like the Principal Agent and Gateway Agents, ensures that distributed resources are optimally utilized, enabling seamless workflow management and global agent cooperation across different organizations. The architecture also enables small and large participants to collaborate across the lifecyle of agentic systems.

The DAWN platform's contributions—hierarchical architecture engendering flexibility, scalability, interoperability, safety, and compliance—position it as a transformative solution for applications of varying complexity. It shows how agents may be registered, discovered, connected and threaded into workflows to meet the needs of agentic software development. DAWN also provides control points to address the safety, security, privacy, and compliance requirements of agentic software development within both the Principal Agent and Gateway Agents. The Principal Agent is wholly owned and operated by the organization that develops it and therefore its make up falls entirely within its purview. Gateway Agents, on the other hand, may be owned operated by subsidiaries, partners, and even competitors. The DAWN architecture therefore provides for maximum flexibility while also allowing for requisite compliance considerations.

The innovative architecture of DAWN inspires future directions in agentic learning with abundant opportunities for agents to learn and adapt based on their interactions within the network. In DAWN, globally distributed agents have the potential to improve their effectiveness and presentation through manifest files, initially crafted by their developers to describe their resources. Gateway Agents play a critical role here, gathering insights on agents' functionality, latency, and security features, and adding this information as metadata to the agents' manifest files. Through repeated interactions, the Principal Agent learns from its experiences with different Gateway Agents, optimizing its query patterns and enhancing its ability to select the most suitable Gateway Agents for specific tasks. In turn, Gateway Agents learn about the agents that are registered with them. As each component gains experience and collaborates within the DAWN framework, they contribute to a collective improvement in efficiency, making task execution more cost-effective, secure, and time-efficient.

The evaluation of complex agentic workflows remains a dynamic field of research, with considerable focus on developing benchmarks and metrics for multi-agent applications. The lack of standardized benchmarks poses challenges, as does the assessment of individual network components in isolation -- such as the Principal Agent -- due to the inherent complexity and non-determinism in multi-agent networks. Establishing clear metrics and success measures for multi-agent systems is essential, and DAWN’s architecture encourages further research and refinement in this field, potentially setting standards for agentic networks of the future.

\section{Acknowledgments}
This research is supported by Outshift AI, the innovation engine of Cisco Systems, Inc. We extend our profound thanks to Reinaldo Penno, Distinguished Engineer, Outshift, for his diligent review of our paper. We also express our sincere appreciation to Nihar Dandekar, Director, Outshift, for sharing insights on potential solutions for agentic frameworks.

\bibliographystyle{IEEEtran}  

\begin{thebibliography}{10}
\providecommand{\url}[1]{#1}
\csname url@samestyle\endcsname
\providecommand{\newblock}{\relax}
\providecommand{\bibinfo}[2]{#2}
\providecommand{\BIBentrySTDinterwordspacing}{\spaceskip=0pt\relax}
\providecommand{\BIBentryALTinterwordstretchfactor}{4}
\providecommand{\BIBentryALTinterwordspacing}{\spaceskip=\fontdimen2\font plus
\BIBentryALTinterwordstretchfactor\fontdimen3\font minus \fontdimen4\font\relax}
\providecommand{\BIBforeignlanguage}[2]{{%
\expandafter\ifx\csname l@#1\endcsname\relax
\typeout{** WARNING: IEEEtran.bst: No hyphenation pattern has been}%
\typeout{** loaded for the language `#1'. Using the pattern for}%
\typeout{** the default language instead.}%
\else
\language=\csname l@#1\endcsname
\fi
#2}}
\providecommand{\BIBdecl}{\relax}
\BIBdecl

\bibitem{openai2024chatgpt}
OpenAI, ``Chatgpt,'' \url{https://www.openai.com/chatgpt}, 2024.

\bibitem{touvron2024llama3}
\BIBentryALTinterwordspacing
H.~Touvron \emph{et~al.}, ``The llama 3 herd of models,'' \emph{arXiv preprint arXiv:2407.21783}, 2024. [Online]. Available: \url{https://arxiv.org/abs/2407.21783}
\BIBentrySTDinterwordspacing

\bibitem{claude3}
Anthropic, ``The claude 3 model family: Opus, sonnet, haiku,'' \url{https://www.anthropic.com/research}, 2024.

\bibitem{guan2024intelligent}
Y.~Guan, D.~Wang, Z.~Chu, S.~Wang, F.~Ni, R.~Song, and C.~Zhuang, ``Intelligent agents with llm-based process automation,'' in \emph{Proceedings of the 30th ACM SIGKDD Conference on Knowledge Discovery and Data Mining (KDD '24)}.\hskip 1em plus 0.5em minus 0.4em\relax ACM, 2024, pp. 5018--5027.

\bibitem{openinterpreter2024}
O.~I. Contributors, ``Open interpreter: A natural language interface for code execution,'' \url{https://github.com/OpenInterpreter/open-interpreter}, 2024, accessed: 2024-09-15.

\bibitem{nijkamp2023codegen}
E.~Nijkamp, B.~Pang, H.~Hayashi, L.~Tu, H.~Wang, Y.~Zhou, S.~Savarese, and C.~Xiong, ``Codegen: An open large language model for code with multi-turn program synthesis,'' in \emph{The Eleventh International Conference on Learning Representations (ICLR)}, 2023.

\bibitem{zhou2024webarena}
S.~Zhou, F.~F. Xu, H.~Zhu, X.~Zhou, R.~Lo, A.~Sridhar, X.~Cheng, T.~Ou, Y.~Bisk, D.~Fried, U.~Alon, and G.~Neubig, ``Webarena: A realistic web environment for building autonomous agents,'' in \emph{The Twelfth International Conference on Learning Representations (ICLR)}, 2024.

\bibitem{wang2024voyager}
G.~Wang, Y.~Xie, Y.~Jiang, A.~Mandlekar, C.~Xiao, Y.~Zhu, L.~Fan, and A.~Anandkumar, ``Voyager: An open-ended embodied agent with large language models,'' \emph{Transactions on Machine Learning Research (TMLR)}, 2024.

\bibitem{zhang2024proagent}
C.~Zhang, K.~Yang, S.~Hu, Z.~Wang, G.~Li, Y.~Sun, C.~Zhang, Z.~Zhang, A.~Liu, S.-C. Zhu \emph{et~al.}, ``Proagent: building proactive cooperative agents with large language models,'' in \emph{Proceedings of the AAAI Conference on Artificial Intelligence}, vol.~38, no.~16, 2024, pp. 17\,591--17\,599.

\bibitem{wuautogen}
Q.~Wu, G.~Bansal, J.~Zhang, Y.~Wu, B.~Li, E.~Zhu, L.~Jiang, X.~Zhang, S.~Zhang, J.~Liu \emph{et~al.}, ``Autogen: Enabling next-gen llm applications via multi-agent conversations,'' in \emph{First Conference on Language Modeling}, 2024.

\bibitem{hong2024metagpt}
S.~Hong, M.~Zhuge, J.~Chen, X.~Zheng, Y.~Cheng, J.~Wang, C.~Zhang, Z.~Wang, S.~K.~S. Yau, Z.~Lin, L.~Zhou, C.~Ran, L.~Xiao, C.~Wu, and J.~Schmidhuber, ``Meta{GPT}: Meta programming for a multi-agent collaborative framework,'' in \emph{The Twelfth International Conference on Learning Representations (ICLR)}, 2024.

\bibitem{li2023camel}
G.~Li, H.~Hammoud, H.~Itani, D.~Khizbullin, and B.~Ghanem, ``Camel: Communicative agents for" mind" exploration of large language model society,'' \emph{Advances in Neural Information Processing Systems (NeurIPS)}, vol.~36, pp. 51\,991--52\,008, 2023.

\bibitem{chen2024agentverse}
W.~Chen, Y.~Su, J.~Zuo, C.~Yang, C.~Yuan, C.-M. Chan, H.~Yu, Y.~Lu, Y.-H. Hung, C.~Qian, Y.~Qin, X.~Cong, R.~Xie, Z.~Liu, M.~Sun, and J.~Zhou, ``Agentverse: Facilitating multi-agent collaboration and exploring emergent behaviors,'' in \emph{The Twelfth International Conference on Learning Representations (ICLR)}, 2024.

\bibitem{langgraph2023}
LangChain, ``Langgraph: A graph-based framework for building ai agents,'' \url{https://www.langchain.com/langgraph}, 2023, accessed: 2024-09-15.

\bibitem{openai_swarm}
\BIBentryALTinterwordspacing
OpenAI, ``Swarm: An openai repository for multi-agent research,'' 2024, accessed: 2024-11-07. [Online]. Available: \url{https://github.com/openai/swarm}
\BIBentrySTDinterwordspacing

\bibitem{magenticone2024}
\BIBentryALTinterwordspacing
{Microsoft Research}, ``{Magentic One}: A generalist multi-agent system for solving complex tasks,'' 2024, accessed: 2024-11-07. [Online]. Available: \url{https://www.microsoft.com/en-us/research/articles/magentic-one-a-generalist-multi-agent-system-for-solving-complex-tasks/}
\BIBentrySTDinterwordspacing

\bibitem{cisco_agntcy_2025}
{Cisco}, ``Agntcy: Internet of agents,'' \url{https://docs.agntcy.org/}, 2025.

\bibitem{google_a2a_2025}
{Google}, ``{Agent2Agent (A2A) Protocol},'' \url{https://github.com/google/A2A}, 2025.

\bibitem{agentcommunicationprotocol_welcome_2025}
{IBM}, ``Welcome to the agent communication protocol,'' \url{https://agentcommunicationprotocol.dev/introduction/welcome}, 2025.

\bibitem{xia2024agentlessdemystifyingllmbasedsoftware}
\BIBentryALTinterwordspacing
C.~S. Xia, Y.~Deng, S.~Dunn, and L.~Zhang, ``Agentless: Demystifying llm-based software engineering agents,'' 2024. [Online]. Available: \url{https://arxiv.org/abs/2407.01489}
\BIBentrySTDinterwordspacing

\bibitem{greshake2023youvesignedforcompromising}
K.~Greshake, S.~Abdelnabi, S.~Mishra, C.~Endres, T.~Holz, and M.~Fritz, ``Not what you've signed up for: Compromising real-world llm-integrated applications with indirect prompt injection,'' in \emph{Proceedings of the 16th ACM Workshop on Artificial Intelligence and Security}, ser. AISec '23, 2023, p. 79–90.

\bibitem{yu2024gptfuzzerredteaminglarge}
\BIBentryALTinterwordspacing
J.~Yu, X.~Lin, Z.~Yu, and X.~Xing, ``Gptfuzzer: Red teaming large language models with auto-generated jailbreak prompts,'' 2024. [Online]. Available: \url{https://arxiv.org/abs/2309.10253}
\BIBentrySTDinterwordspacing

\bibitem{owasp2025llm}
{OWASP GenAI Security Project}, ``{OWASP Top 10 for Large Language Model Applications 2025},'' \url{https://genai.owasp.org/resource/owasp-top-10-for-llm-applications-2025/}, 2025.

\bibitem{openai2024}
\BIBentryALTinterwordspacing
OpenAI, ``Introducing the gpt store,'' 2024, accessed: 2024-11-07. [Online]. Available: \url{https://openai.com/index/introducing-the-gpt-store/}
\BIBentrySTDinterwordspacing

\bibitem{zheng2023judging}
L.~Zheng, W.-L. Chiang, Y.~Sheng, S.~Zhuang, Z.~Wu, Y.~Zhuang, Z.~Lin, Z.~Li, D.~Li, E.~Xing \emph{et~al.}, ``Judging llm-as-a-judge with mt-bench and chatbot arena,'' \emph{Advances in Neural Information Processing Systems}, vol.~36, pp. 46\,595--46\,623, 2023.

\bibitem{wooldridge1995intelligent}
M.~Wooldridge and N.~R. Jennings, ``Intelligent agents: Theory and practice,'' \emph{The Knowledge Engineering Review}, vol.~10, no.~2, pp. 115--152, 1995.

\bibitem{wang2024survey}
L.~Wang, C.~Ma, X.~Feng, Z.~Zhang, H.~Yang, J.~Zhang, Z.~Chen, J.~Tang, X.~Chen, Y.~Lin \emph{et~al.}, ``A survey on large language model based autonomous agents,'' \emph{Frontiers of Computer Science}, vol.~18, no.~6, p. 186345, 2024.

\bibitem{wei2022chain}
J.~Wei, X.~Wang, D.~Schuurmans, M.~Bosma, F.~Xia, E.~Chi, Q.~V. Le, D.~Zhou \emph{et~al.}, ``Chain-of-thought prompting elicits reasoning in large language models,'' \emph{Advances in neural information processing systems}, vol.~35, pp. 24\,824--24\,837, 2022.

\bibitem{yao2022react}
\BIBentryALTinterwordspacing
S.~Yao, D.~Yu, J.~Z. Lu, I.~Shafran, K.~Narasimhan, Y.~Cao, and W.~Chen, ``React: Synergizing reasoning and acting in language models,'' \emph{arXiv preprint arXiv:2210.03629}, 2022. [Online]. Available: \url{https://arxiv.org/abs/2210.03629}
\BIBentrySTDinterwordspacing

\bibitem{yao2024tree}
S.~Yao, D.~Yu, J.~Zhao, I.~Shafran, T.~Griffiths, Y.~Cao, and K.~Narasimhan, ``Tree of thoughts: Deliberate problem solving with large language models,'' \emph{Advances in Neural Information Processing Systems (NeurIPS)}, vol.~36, 2024.

\bibitem{salas2000teamwork}
\BIBentryALTinterwordspacing
E.~Salas, K.~C. Stagl, C.~S. Burke, and G.~F. Goodwin, ``Teamwork: Emerging principles,'' \emph{Philosophy and Technology}, vol.~7, no.~3, pp. 341--350, 2000. [Online]. Available: \url{https://onlinelibrary.wiley.com/doi/abs/10.1111/1468-2370.00046}
\BIBentrySTDinterwordspacing

\bibitem{qian2023chatdev}
\BIBentryALTinterwordspacing
C.~Qian, W.~Liu, H.~Liu, N.~Chen, Y.~Dang, J.~Li, C.~Yang, W.~Chen, Y.~Su, X.~Cong, J.~Xu, D.~Li, Z.~Liu, and M.~Sun, ``Chatdev: Communicative agents for software development,'' in \emph{ACL (1)}, 2024, pp. 15\,174--15\,186. [Online]. Available: \url{https://aclanthology.org/2024.acl-long.810}
\BIBentrySTDinterwordspacing

\bibitem{liu2023agentcoder}
\BIBentryALTinterwordspacing
Y.~Liu, H.~Zhang, W.~Wang, and et~al., ``Agentcoder: Multi-agent-based code generation with iterative testing and optimisation,'' \emph{arXiv preprint arXiv:2308.09351}, 2023. [Online]. Available: \url{https://arxiv.org/abs/2308.09351}
\BIBentrySTDinterwordspacing

\bibitem{lin2023medagents}
\BIBentryALTinterwordspacing
X.~Tang, A.~Zou, Z.~Zhang, Z.~Li, Y.~Zhao, X.~Zhang, A.~Cohan, and M.~Gerstein, ``Medagents: Large language models as collaborators for zero-shot medical reasoning,'' in \emph{Findings of the Association for Computational Linguistics: ACL 2024}.\hskip 1em plus 0.5em minus 0.4em\relax Association for Computational Linguistics, 2024, pp. 599--621. [Online]. Available: \url{https://aclanthology.org/2024.findings-acl.33}
\BIBentrySTDinterwordspacing

\bibitem{ju2024floodingspreadmanipulatedknowledge}
\BIBentryALTinterwordspacing
T.~Ju, Y.~Wang, X.~Ma, P.~Cheng, H.~Zhao, Y.~Wang, L.~Liu, J.~Xie, Z.~Zhang, and G.~Liu, ``Flooding spread of manipulated knowledge in llm-based multi-agent communities,'' 2024. [Online]. Available: \url{https://arxiv.org/abs/2407.07791}
\BIBentrySTDinterwordspacing

\bibitem{chen2024ioa}
W.~Chen, Z.~You, R.~Li, Y.~Guan, C.~Qian, C.~Zhao, C.~Yang, R.~Xie, Z.~Liu, and M.~Sun, ``Internet of agents: Weaving a web of heterogeneous agents for collaborative intelligence,'' \emph{arXiv preprint arXiv:2407.07061}, 2024.

\bibitem{xu2023rewoodecouplingreasoningobservations}
\BIBentryALTinterwordspacing
B.~Xu, Z.~Peng, B.~Lei, S.~Mukherjee, Y.~Liu, and D.~Xu, ``Rewoo: Decoupling reasoning from observations for efficient augmented language models,'' 2023. [Online]. Available: \url{https://arxiv.org/abs/2305.18323}
\BIBentrySTDinterwordspacing

\bibitem{shen2024hugginggpt}
Y.~Shen, K.~Song, X.~Tan, D.~Li, W.~Lu, and Y.~Zhuang, ``Hugginggpt: Solving ai tasks with chatgpt and its friends in hugging face,'' \emph{Advances in Neural Information Processing Systems (NeurIPS)}, vol.~36, 2024.

\bibitem{huggingface}
\BIBentryALTinterwordspacing
HuggingFace, ``Hugging face: The ai community building the future.'' [Online]. Available: \url{https://huggingface.co}
\BIBentrySTDinterwordspacing

\bibitem{zhou2024languageagenttreesearch}
\BIBentryALTinterwordspacing
A.~Zhou, K.~Yan, M.~Shlapentokh-Rothman, H.~Wang, and Y.-X. Wang, ``Language agent tree search unifies reasoning acting and planning in language models,'' 2024. [Online]. Available: \url{https://arxiv.org/abs/2310.04406}
\BIBentrySTDinterwordspacing

\bibitem{qin2024toolllm}
Y.~Qin, S.~Liang, Y.~Ye, K.~Zhu, L.~Yan, Y.~Lu, Y.~Lin, X.~Cong, X.~Tang, B.~Qian, S.~Zhao, L.~Hong, R.~Tian, R.~Xie, J.~Zhou, M.~Gerstein, dahai li, Z.~Liu, and M.~Sun, ``Tool{LLM}: Facilitating large language models to master 16000+ real-world {API}s,'' in \emph{The Twelfth International Conference on Learning Representations (ICLR)}, 2024.

\bibitem{ALFWorld20}
\BIBentryALTinterwordspacing
M.~Shridhar, X.~Yuan, M.-A. C\^ot\'e, Y.~Bisk, A.~Trischler, and M.~Hausknecht, ``{ALFWorld: Aligning Text and Embodied Environments for Interactive Learning},'' in \emph{Proceedings of the International Conference on Learning Representations (ICLR)}, 2021. [Online]. Available: \url{https://arxiv.org/abs/2010.03768}
\BIBentrySTDinterwordspacing

\bibitem{abuelsaad2024agenteautonomouswebnavigation}
\BIBentryALTinterwordspacing
T.~Abuelsaad, D.~Akkil, P.~Dey, A.~Jagmohan, A.~Vempaty, and R.~Kokku, ``Agent-e: From autonomous web navigation to foundational design principles in agentic systems,'' 2024. [Online]. Available: \url{https://arxiv.org/abs/2407.13032}
\BIBentrySTDinterwordspacing

\bibitem{openaifunctioncalling}
\BIBentryALTinterwordspacing
OpenAI. Openai guides on function calling. [Online]. Available: \url{https://platform.openai.com/docs/guides/function-calling}
\BIBentrySTDinterwordspacing

\bibitem{modelcontextprotocol_intro_2025}
{Model Context Protocol}, ``Introduction to the model context protocol,'' \url{https://modelcontextprotocol.io/introduction}, 2025.

\bibitem{kim2024leveraging}
M.~Kim, T.~Stennett, D.~Shah, S.~Sinha, and A.~Orso, ``Leveraging large language models to improve rest api testing,'' in \emph{Proceedings of the 2024 ACM/IEEE 44th International Conference on Software Engineering: New Ideas and Emerging Results}, 2024, pp. 37--41.

\bibitem{rapidapi_api_testing}
\BIBentryALTinterwordspacing
RapidAPI, ``Api testing guide: Everything you need to know,'' RapidAPI Blog. [Online]. Available: \url{https://rapidapi.com/blog/api-testing/}
\BIBentrySTDinterwordspacing

\bibitem{apache_jmeter}
\BIBentryALTinterwordspacing
{The Apache Software Foundation}, \emph{Apache JMeter}, 2024, accessed: 2024-09-25. [Online]. Available: \url{https://jmeter.apache.org/}
\BIBentrySTDinterwordspacing

\bibitem{gatling_tool}
\BIBentryALTinterwordspacing
GatlingCorp, \emph{Gatling Open-Source Load Testing Tool}, 2024, accessed: 2024-09-25. [Online]. Available: \url{https://gatling.io/}
\BIBentrySTDinterwordspacing

\bibitem{postman}
\BIBentryALTinterwordspacing
I.~Postman, \emph{Postman API Platform}, 2024, accessed: 2024-09-25. [Online]. Available: \url{https://www.postman.com/}
\BIBentrySTDinterwordspacing

\bibitem{manning2009introduction}
C.~D. Manning, \emph{An introduction to information retrieval}.\hskip 1em plus 0.5em minus 0.4em\relax Cambridge University Press, 2009.

\bibitem{stoica2003semantic}
\BIBentryALTinterwordspacing
I.~Stoica, R.~Morris, D.~R. Karger, M.~F. Kaashoek, and H.~Balakrishnan, ``Semantic search,'' in \emph{Proceedings of the 12th international conference on World Wide Web}.\hskip 1em plus 0.5em minus 0.4em\relax ACM, 2003, pp. 700--709. [Online]. Available: \url{https://dl.acm.org/doi/abs/10.1145/775152.775250}
\BIBentrySTDinterwordspacing

\bibitem{patil2024gorilla}
S.~G. Patil, T.~Zhang, X.~Wang, and J.~E. Gonzalez, ``Gorilla: Large language model connected with massive {API}s,'' in \emph{The Thirty-eighth Annual Conference on Neural Information Processing Systems (NeurIPS)}, 2024.

\bibitem{granjal2010secure}
J.~Granjal, E.~Monteiro, and J.~S.~S. Silva, ``A secure interconnection model for ipv6 enabled wireless sensor networks,'' in \emph{2010 IFIP Wireless Days}.\hskip 1em plus 0.5em minus 0.4em\relax IEEE, 2010, pp. 1--6.

\bibitem{jebri2015efficient}
S.~Jebri, M.~Abid, and A.~Bouallegue, ``An efficient scheme for anonymous communication in iot,'' in \emph{2015 11th International Conference on Information Assurance and Security (IAS)}.\hskip 1em plus 0.5em minus 0.4em\relax IEEE, 2015, pp. 7--12.

\bibitem{jing2014security}
Q.~Jing, A.~V. Vasilakos, J.~Wan, J.~Lu, and D.~Qiu, ``Security of the internet of things: perspectives and challenges,'' \emph{Wireless networks}, vol.~20, pp. 2481--2501, 2014.

\bibitem{shavit2023practices}
Y.~Shavit, S.~Agarwal, M.~Brundage, S.~Adler, C.~O’Keefe, R.~Campbell, T.~Lee, P.~Mishkin, T.~Eloundou, A.~Hickey \emph{et~al.}, ``Practices for governing agentic ai systems,'' \emph{Research Paper, OpenAI}, 2023.

\bibitem{huang2023survey}
L.~Huang, W.~Yu, W.~Ma, W.~Zhong, Z.~Feng, H.~Wang, Q.~Chen, W.~Peng, X.~Feng, B.~Qin \emph{et~al.}, ``A survey on hallucination in large language models: Principles, taxonomy, challenges, and open questions,'' \emph{arXiv preprint arXiv:2311.05232}, 2023.

\bibitem{li2023multistep}
H.~Li, D.~Guo, W.~Fan, M.~Xu, J.~Huang, F.~Meng, and Y.~Song, ``Multi-step jailbreaking privacy attacks on chat{GPT},'' in \emph{The 2023 Conference on Empirical Methods in Natural Language Processing (EMNLP)}, 2023.

\bibitem{shen2024donowcharacterizingevaluating}
\BIBentryALTinterwordspacing
X.~Shen, Z.~Chen, M.~Backes, Y.~Shen, and Y.~Zhang, ``"do anything now": Characterizing and evaluating in-the-wild jailbreak prompts on large language models,'' 2024. [Online]. Available: \url{https://arxiv.org/abs/2308.03825}
\BIBentrySTDinterwordspacing

\bibitem{10.1145/3605764.3623985}
\BIBentryALTinterwordspacing
K.~Greshake, S.~Abdelnabi, S.~Mishra, C.~Endres, T.~Holz, and M.~Fritz, ``Not what you've signed up for: Compromising real-world llm-integrated applications with indirect prompt injection,'' in \emph{Proceedings of the 16th ACM Workshop on Artificial Intelligence and Security}, ser. AISec '23, 2023, p. 79–90. [Online]. Available: \url{https://doi.org/10.1145/3605764.3623985}
\BIBentrySTDinterwordspacing

\bibitem{zou2024poisonedragknowledgecorruptionattacks}
\BIBentryALTinterwordspacing
W.~Zou, R.~Geng, B.~Wang, and J.~Jia, ``Poisonedrag: Knowledge corruption attacks to retrieval-augmented generation of large language models,'' 2024. [Online]. Available: \url{https://arxiv.org/abs/2402.07867}
\BIBentrySTDinterwordspacing

\bibitem{zhan-etal-2024-injecagent}
\BIBentryALTinterwordspacing
Q.~Zhan, Z.~Liang, Z.~Ying, and D.~Kang, ``{I}njec{A}gent: Benchmarking indirect prompt injections in tool-integrated large language model agents,'' in \emph{Findings of the Association for Computational Linguistics: ACL 2024}, 2024, pp. 10\,471--10\,506. [Online]. Available: \url{https://aclanthology.org/2024.findings-acl.624}
\BIBentrySTDinterwordspacing

\bibitem{he2024emerged}
F.~He, T.~Zhu, D.~Ye, B.~Liu, W.~Zhou, and P.~S. Yu, ``The emerged security and privacy of llm agent: A survey with case studies,'' \emph{arXiv preprint arXiv:2407.19354}, 2024.

\bibitem{zhang2024s}
Z.~Zhang, M.~Jia, H.-P. Lee, B.~Yao, S.~Das, A.~Lerner, D.~Wang, and T.~Li, ``“it's a fair game”, or is it? examining how users navigate disclosure risks and benefits when using llm-based conversational agents,'' in \emph{Proceedings of the CHI Conference on Human Factors in Computing Systems}, 2024, pp. 1--26.

\bibitem{ruan2024identifying}
Y.~Ruan, H.~Dong, A.~Wang, S.~Pitis, Y.~Zhou, J.~Ba, Y.~Dubois, C.~J. Maddison, and T.~Hashimoto, ``Identifying the risks of {LM} agents with an {LM}-emulated sandbox,'' in \emph{The Twelfth International Conference on Learning Representations (ICLR)}, 2024.

\bibitem{welbl2021challengesdetoxifyinglanguagemodels}
\BIBentryALTinterwordspacing
J.~Welbl, A.~Glaese, J.~Uesato, S.~Dathathri, J.~Mellor, L.~A. Hendricks, K.~Anderson, P.~Kohli, B.~Coppin, and P.-S. Huang, ``Challenges in detoxifying language models,'' in \emph{Findings of the Association for Computational Linguistics: EMNLP 2021}, 2021, pp. 2447--2469. [Online]. Available: \url{https://aclanthology.org/2021.findings-emnlp.210}
\BIBentrySTDinterwordspacing

\bibitem{gehman2020realtoxicitypromptsevaluatingneuraltoxic}
\BIBentryALTinterwordspacing
S.~Gehman, S.~Gururangan, M.~Sap, Y.~Choi, and N.~A. Smith, ``{R}eal{T}oxicity{P}rompts: Evaluating neural toxic degeneration in language models,'' in \emph{Findings of the Association for Computational Linguistics: EMNLP 2020}, 2020, pp. 3356--3369. [Online]. Available: \url{https://aclanthology.org/2020.findings-emnlp.301}
\BIBentrySTDinterwordspacing

\bibitem{patel2025building}
\BIBentryALTinterwordspacing
A.~Patel, S.~Shah, S.~Shroff, K.~Athrey, M.~Viswanathan, and R.~Penno, ``Building distributed multi-framework, multi-agent solutions,'' Outshift blog, Cisco Systems, Inc., Apr. 2025. [Online]. Available: \url{https://outshift.cisco.com/blog/building-multi-framework-multi-agent-solutions}
\BIBentrySTDinterwordspacing

\end{thebibliography}

\end{document}